# Sculpting of Martian brain terrain reveals the drying of ancient Mars

**Shenyi Zhang**[1,2], **Lei Zhang**[1,*], **Yutian Ke**[3], **Jinhai Zhang**[1]

[1] *Institute of Geology and Geophysics, Chinese Academy of Sciences, Beijing, China*

[2] *College of Earth and Planetary Sciences, University of Chinese Academy of Sciences, Beijing, China*

[3] *Division of Geological and Planetary Sciences, California Institute of Technology, Pasadena, CA, USA*

**Corresponding authors:** Lei Zhang

**Email:** zhangl@mail.iggcas.ac.cn

## ABSTRACT

The Martian brain terrain (MBT), characterized by its unique brain-like morphology, is a potential geological archive for finding hints of paleoclimatic conditions during its formation period. The morphological similarity of MBT to self-organized patterned ground on Earth suggests a shared formation mechanism. However, the lack of quantitative descriptions and robust physical modeling of self-organized stone transport jointly limits the study of the thermal and aqueous conditions governing MBT's formation. Here we established a specialized quantitative system for extracting the morphological features of MBT, taking a typical region located in the northern Arabia Terra as an example, and then employed a numerical model to investigate its formation mechanisms. Our simulation results accurately replicate the observed morphology of MBT, matching its key geometric metrics with deviations <15%. Crucially, however, we find that the self-organized transport can solely produce relief <0.5 m, insufficient to explain the formation of MBT with average relief of 3.29 ± 0.65 m. We attribute this discrepancy to sculpting driven by late-stage sublimation, constraining cumulative subsurface ice loss in this region to ~3 meters over the past ~3Ma. These findings demonstrate that MBT's formation is a multi-stage process: initial patterning driven by freeze-thaw cycles implying liquid water followed by vertical sculpting via sublimation requiring a dry environment. This evolution provides physical evidence for the transition of the ancient Martian climate from a wetter period to a colder hyper-arid state.

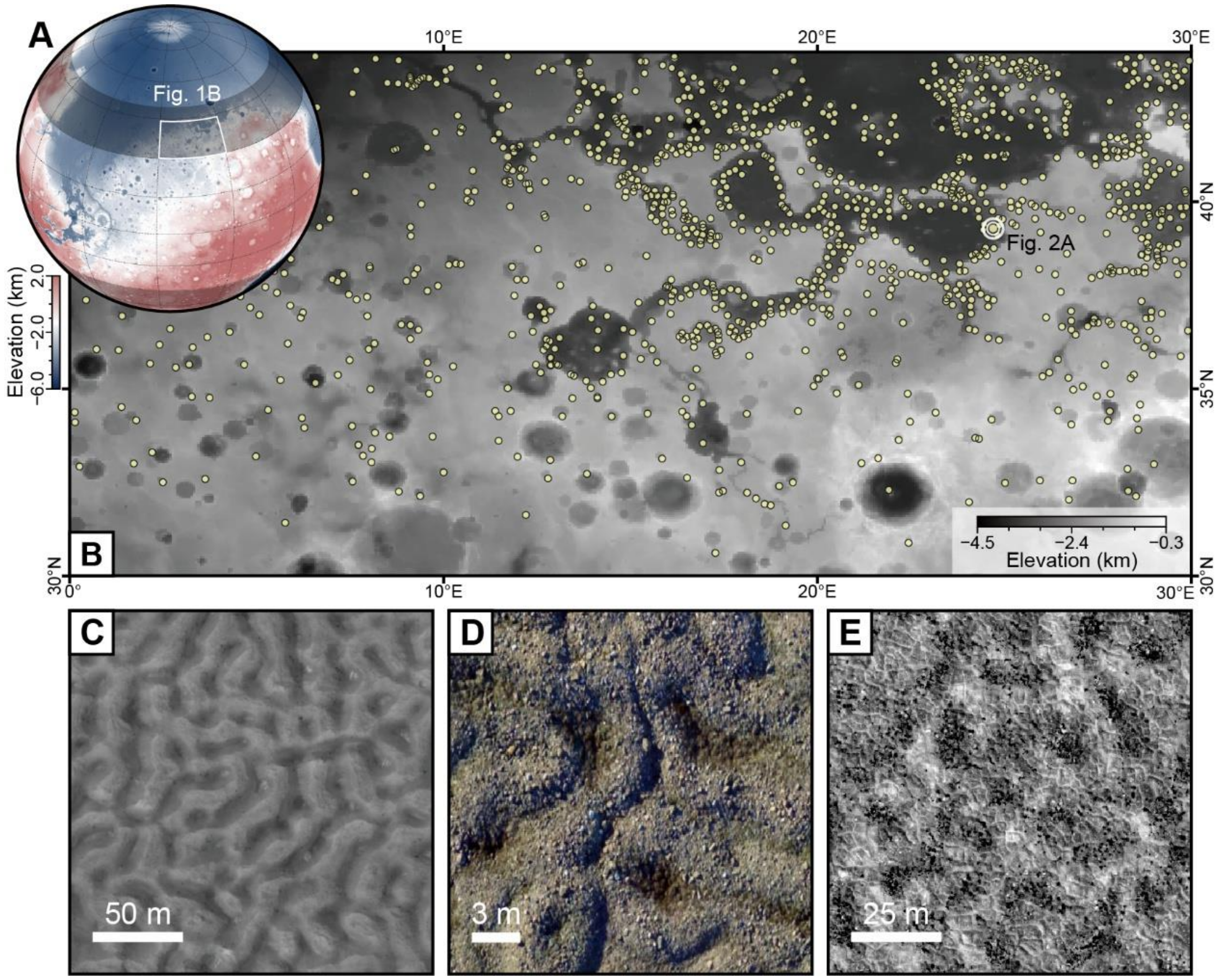

**Figure 1. Martian brain terrain (MBT) distribution and analog**. (A) Study area location on MOLA elevation base (Smith et al., 2001); gray bands show MBT latitude range (Pearson et al., 2024). (B) MBT distribution in the study area. (C) Brain terrain in HiRISE image (ESP_033165_2195). (D) Pattern ground formed by self-organized stone transport, Dundas Harbor, Earth (Hibbard et al., 2020). (E) Densely distributed clusters of stones on Mars (ESP_026770_2525).

## INTRODUCTION

Martian brain terrain (MBT) is a potential archive of Martian paleo-climatic conditions found in the mid-latitudes of both the northern and southern hemispheres (Levy et al., 2009; Cheng et al., 2021; Bina and Osinski, 2021; Pearson et al., 2024; Morgan et al., 2025) (Figs. 1A and B). The close association with glacial landforms (Levy et al., 2010; Souness et al., 2012; Hibbard et al., 2020;

Gallagher et al., 2021; Bina and Osinski, 2021; Driver et al., 2024) and the symmetrical distribution of MBT (Hauber et al., 2008; Levy et al., 2014; Cheng et al., 2021; Pearson et al., 2024) indicate that MBT's formation is controlled by ice/water activities and global climate conditions. These possible processes shaped the MBT's unique morphological pattern, which closely resembles the folds and ridges of human brain (Fig. 1C). Therefore, the distinct morphological feature of MBT is central to understanding its formation mechanism and ultimately provides key insights into the ice/water activities as well as the ancient climate conditions on Mars.

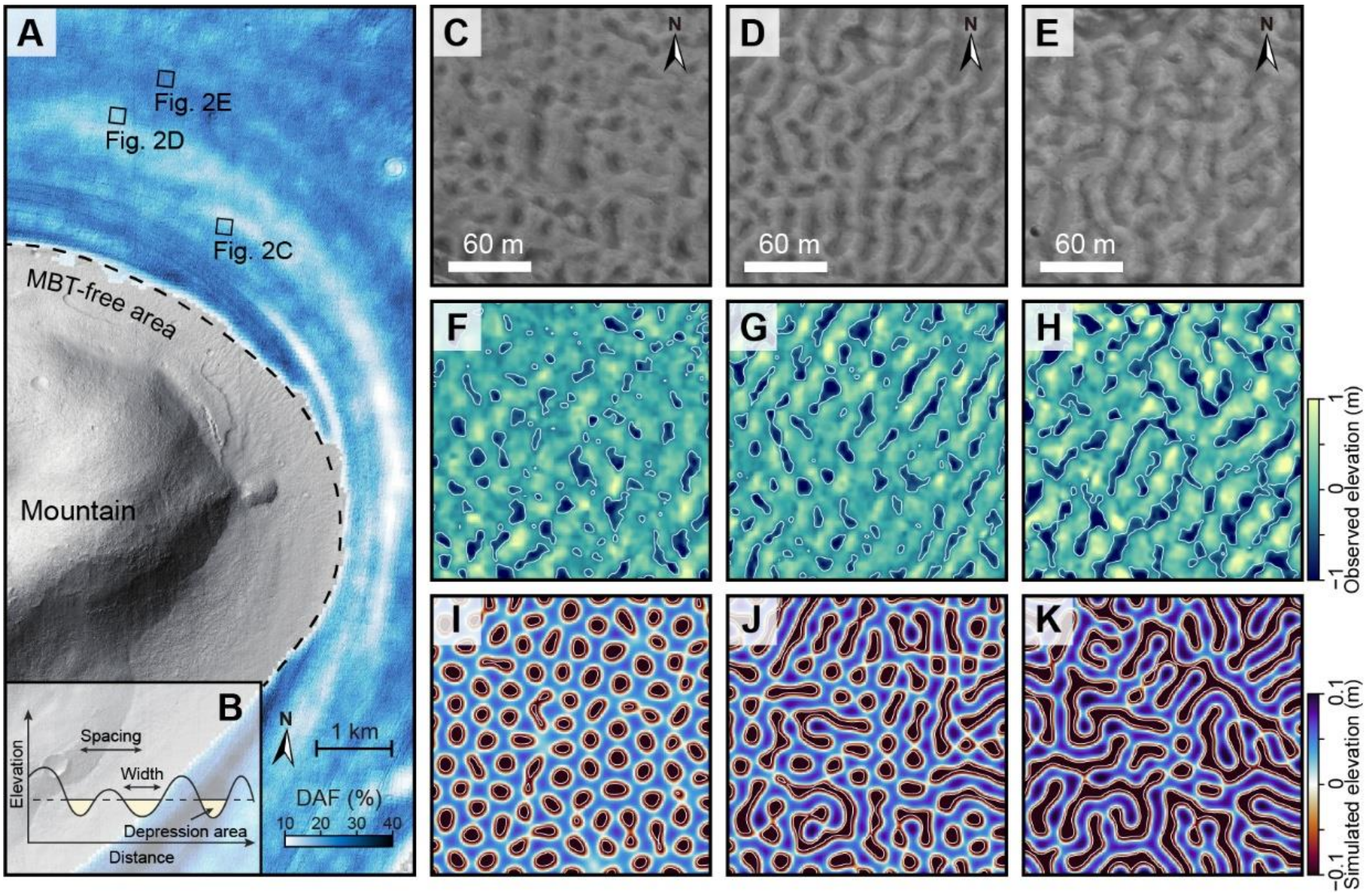

**Figure 2. Morphological metrics and numerical simulations**. (A) Spatial distribution of depression area fraction (DAF) on shaded relief; black dashed line shows the MBT boundary. (B) Schematic definition of key MBT morphological features. The terrain tangent curve, depression areas (yellow), and depression threshold elevation (DTE, dashed line) are shown. (C-E) HiRISE images of three MBT area. (F-H) Detrended elevations for areas in (C-E); white contours outline the depression areas. (I–K) Simulation results at increasing maximum stone movement rates ($v_{max}$). White contours outline depressions.

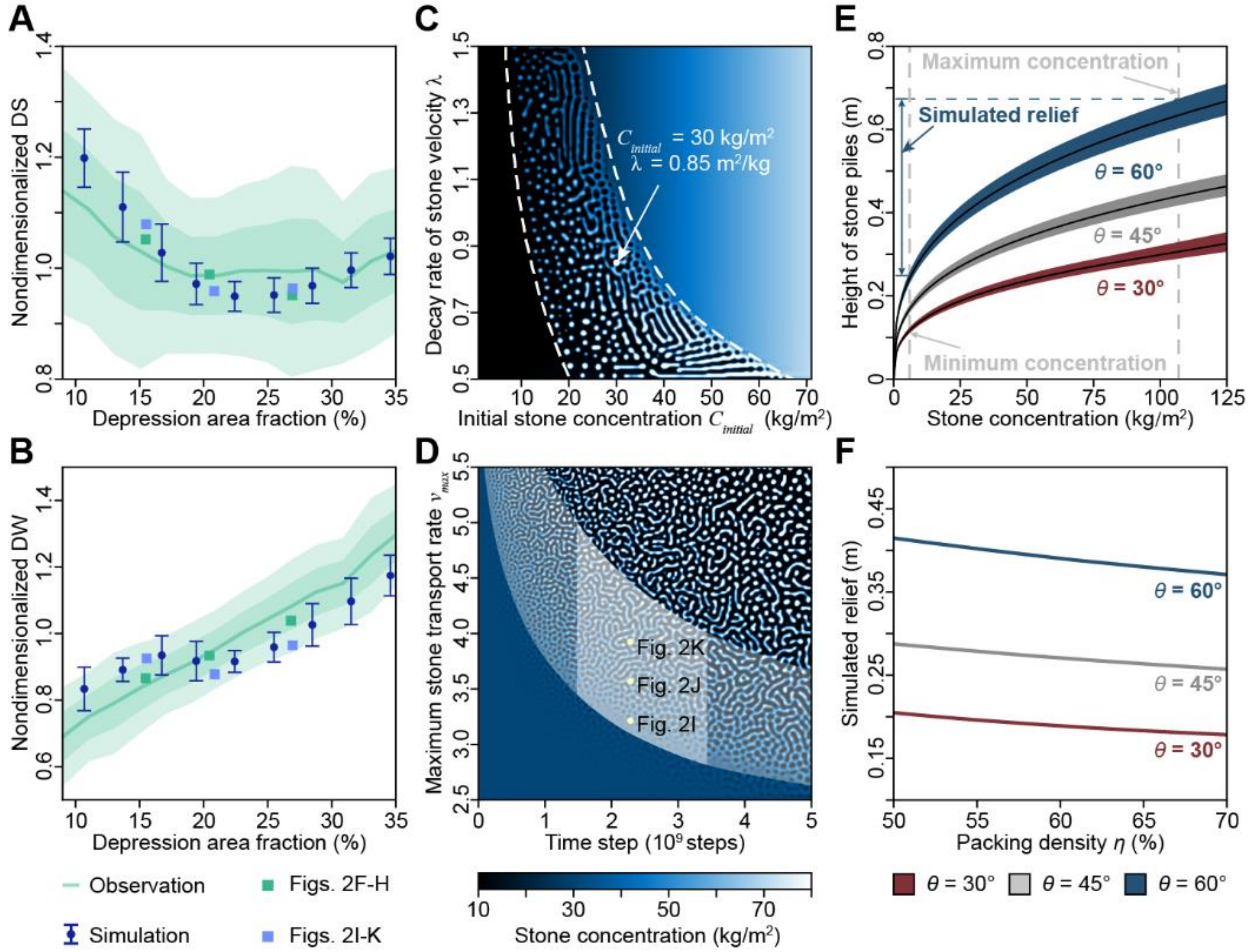

**Figure 3. Comparison between simulated and observed MBT morphology, with parameter scan results and relief constraints.** (A-B) The extraction results for DS and DW, respectively. Light green line and shaded bands show mean ±1σ and ±2σ for observed MBTs. Dark blue scatter points with error bars (SD) are simulation results. Green and dark blue squares correspond to examples in (F−H) and (I−K), respectively. (C) Parameter scan results for $\lambda$ and $c_{initial}$. Dashed line marks the phase separation boundary. The white dot represents the parameter used in this study ($v_{max}$ = 3.0 mm/cycle. $M$ = 1×10$^{-6}$, $d$ = 2, $\alpha$ = 1, $\kappa$ = 0.225, $t$ = 1×10$^{10}$ steps). (D) Parameter scan results for $v_{max}$ and $t$. The three dots represent the parameter used in Figs. 2I-K. ($t$ = 2.3×10$^{9}$ steps, $v_{max}$ = 3.25, 3.60, 3.95 mm/cycle, respectively). The shaded area represents the parameter range used in this study. (E) Theoretical stone pile height vs. stone concentration for different packing densities ($\eta$ = 50−70%) and angles of repose ($\theta$ = 30°, 45°, 60°). Solid black line: $\eta$ = 60%. (F) The theoretical maximum height versus $\eta$. The three curves correspond to $\theta$ of 60°, 45°, and 30°, respectively.

Three formation mechanisms for MBT have been proposed: thermal processes (Mangold, 2003; Levy et al., 2009), aeolian processes (Cheng et al., 2021), and self-organization (Kessler and Werner, 2003; Li et al., 2021). Although the first two mechanisms can account for the topographic relief, they struggle to explain the brain-like geometry. In contrast, self-organization provides a reasonable explanation for the complex patterning. Self-organization is a widespread phenomenon in nature (Liu et al., 2013; Li et al., 2021). On Earth, patterned ground formed by freeze-thaw cycles is a classic example of self-organization (Fig. 1D) and bears a strong morphological resemblance to MBT (Fig. 1E), suggesting that MBT may also be a self-organized landform. However, this interpretation lacks quantitative support due to the difficulty of characterizing sinuous MBT morphology and the absence of mechanistic research.

Here we establish a specialized quantification system for characterizing the morphological features of MBT and develop numerical models to investigate its formation mechanism, clarifying the relationship between the MBT's morphological features and the paleoclimatic conditions and ice/water activities during its formation period.

## RESULTS

### Morphological Parameter Extraction

Following the mapping of MBT distribution in northern Arabia Terra on Mars (Fig. 1B), we selected a representative region (Fig. 2A) to derive quantitative morphological parameters of MBT. Based on the high-resolution digital terrain model (Smith et al., 2001), we conducted extraction of specific morphological parameters—depression area fraction (DAF), depression spacing (DS) and depression width (DW) (Fig. 2A). The extraction results of the study area show DAF ranging from ~7% to ~37%, with average DS and DW of 27.52 ± 2.38 m and 7.78 ± 0.93 m, respectively. Further analysis shows that: The DS initially decreases and then increases with increasing DAF (Fig. 3A). In addition, DW shows a consistent increase with increasing DAF (Fig. 3B).

**The Numerical Simulation of Self-Organized Stone Transport**

Self-organization is a common phenomenon in nature, with phase separation serving as a key physical principle for its description (Li et al., 2021). It has been employed in simulation studies of landforms on Earth analogue to MBT (Kessler and Werner, 2003; Li et al., 2021), proving that the distribution of stones exhibits characteristics of self-organization under freeze-thaw cycles. Here, based on the framework of self-organization, we develop a dynamic model to describe the spatial variations in stone concentration during the self-organized transport process. We performed a parametric scan of $\lambda$, $c_{initial}$, $v_{max}$ and $t$, identified the parameter range over which “brain-pattern” occurs (Figs. 3C and D), and successfully reproduced patterns highly similar to observed MBT (Figs. 2F-K).

To quantitatively evaluate the accuracy of simulated results, we established a quantitative relationship between concentration and elevation (Figs. 3E) to map the simulated stone concentrations to elevation, enabling the extraction of quantitative morphological parameters. The results show close agreement: the deviation in the extracted morphological features between the simulation and the observed MBT results is less than 15%, with the DAF values obtained from the simulation results (15.52%, 20.83%, 26.96%; Figs. 2I-K) agreeing well with those from the observed MBT (15.47%, 20.46%, 26.82%; Figs. 2F-H), respectively. This close agreement extends to the DS and DW parameters. Besides these specific cases, simulated results obtained with $v_{max}$ values ranging from about 3.0 to 5.0 mm/cycle remain consistent with the observed MBT results (Figs. 3A and B). However, it should be noted that the relief of the simulated results are significantly less than that of MBT in study area.

We tested the relationship between concentration and elevation under varying angles of repose ($\theta$) and packing density ($\eta$) to discuss the maximum achievable relief in the simulation results (Figs. 3E-F). The calculation results show that the maximum relief does not exceed 0.5 m across all tested conditions (Fig. 3F). However, the observation of MBT in the study area reveals an average relief of 3.29 ± 0.65 m, which is obviously larger than 0.5 m. This discrepancy suggests that the formation of the MBT is complex and cannot be attributed solely to the self-organized transport of stones.

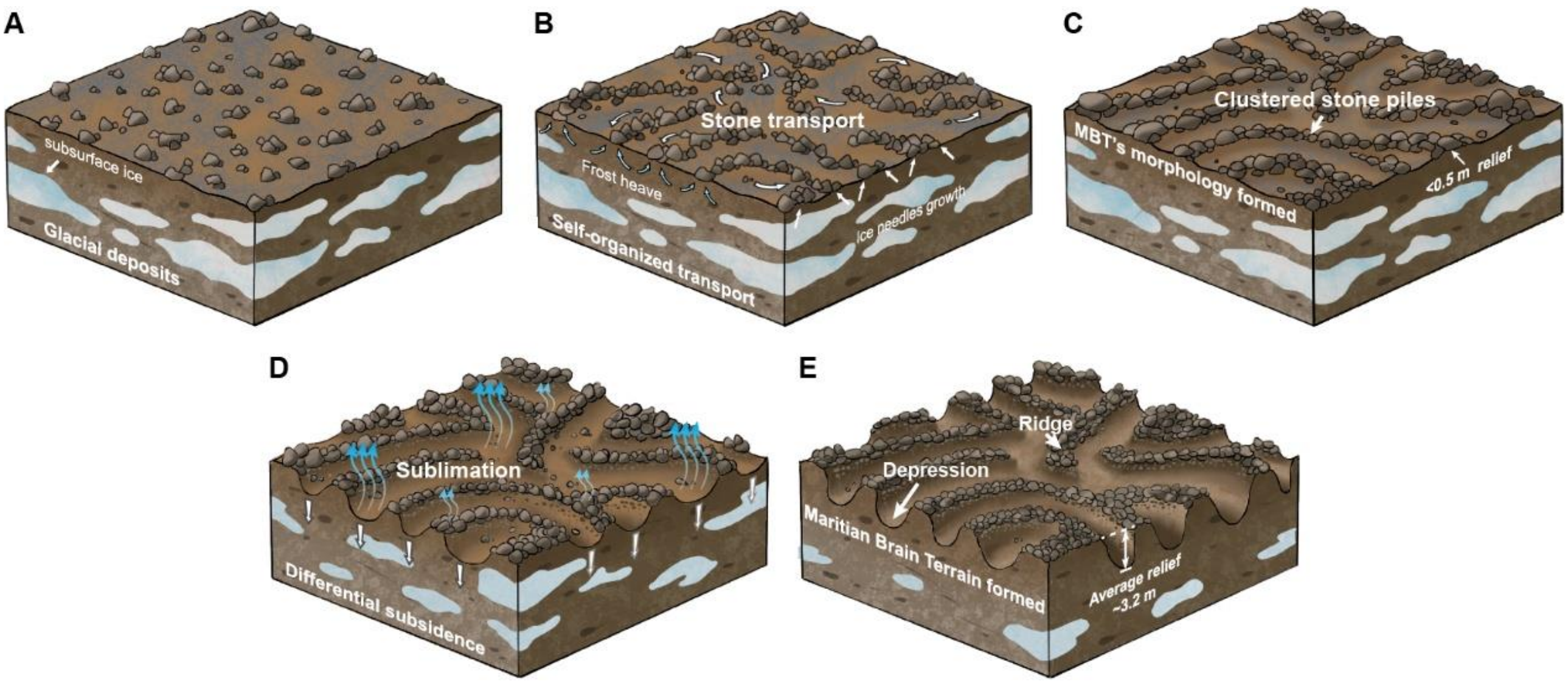

**Figure 4. The two-stage formation mechanism of Martian Brain Terrain.** (A) Initial stage: uniform stone distribution. (B) Self-organized stone transport driven by freeze-thaw cycles; white arrows on top surface indicate the stone transport. The blue arrows on the left-side represent the frost heaving process while white arrows on the right-side indicate the ice needles growth. (C) The brain-like framework (clustered stone piles) formed: raised stone piles (high stone concentration) and depressions (low concentration). (D) The differential sublimation effect enhanced the relief. The light blue arrows represent the sublimation process and the white arrows indicate the subsidence driven by the sublimation process. (E) The Martian brain terrain formed.

## DISCUSSION

### The Implication for the Discrepancy between Numerical Simulation and Observation

The discrepancy in relief between simulated and observed MBT (with <0.5 m and 3.29 ± 0.65 m, respectively) indicates a meter-scale modification process following the self-organization stage. The internal dynamic geological processes (e.g., tectonism and volcanism) fail to explain the modification because they typically reshape terrain on a kilometer scale and leave distinct geological signatures like faults systems and volcano, which are not observed in the study area. Among external dynamic geological processes, impact cratering is localized and random in distribution, while aeolian erosion tends to flatten the surface and cannot account for the regular, deeply incised depression. Sublimation is widespread on Mars and is known to shape mid-latitude glacial landforms (Mangold, 2003; Levy et al., 2009; Zhang et al., 2023). Importantly, it is consistent with the presence of subsurface ice (derived from prior freeze-thaw cycles), operates under thermal conditions distinct from those of freeze-thaw cycles (supporting later modification), and can act uniformly across densely and continuously distributed MBTs. Thus, sublimation is identified as the most likely process resulting in the late-stage modification of the MBT.

### The Implication for the Multi-stage Formation Mechanism of MBT

To systematically account for the vertical discrepancy between simulations and observations, we propose a multi-stage formation mechanism for the MBT that involves an initial self-organization stage followed by a stage dominated by ice sublimation (Fig. 4). During the self-organization stage, repeated freeze-thaw cycles drive stone transport through moisture migration, frost heave, and needle ice growth (Mackay and Mathews, 1974; Berthling et al., 2001; Matsuoka et al., 2003; Yamagishi and Matsuoka, 2015; Li et al., 2018, 2021; Wang et al., 2022; Matsuoka, 2023). Differential displacement—where stones in high-concentration areas move less than those in low-concentration areas—creates positive feedback that concentrates stones further (Li et al., 2021), while spatial and gravitational constraints impose negative feedback. The interplay of these feedbacks progressively organizes stone distribution from random and disordered to regular and orderly. After the self-organization stage, stone clusters create an initial topographic relief. In areas with sparse stone cover (i.e., depressions), ice-rich deposits are largely exposed and receive direct solar radiation. Additionally, owing to the effect of radiative convergence (Moore et al., 2017), sublimation rates in these depressions exceed those in stone-rich areas where the ice-bearing substrate remains covered. This differential sublimation process amplifies the vertical relief. This multi-stage mechanism reflects a transition from a warmer, wetter paleoclimate dominated by freeze-thaw cycles to a colder, drier climate dominated by sublimation, positioning the MBT as a geological archive of Martian paleoclimatic change.

**Timing and Magnitude of Sublimation Modification**

Crater dating specifically targeting MBT within the Ismenius Lacus (including our study area) indicates that MBT evolution was actively ongoing at ~23.1 ± 1.4 Ma and ceased by ~2.9 ± 0.3 Ma (Morgan et al., 2025). These ages correspond to a high obliquity period (~20 Ma) and the period of high-to-low obliquity transition (~5-3 Ma), respectively (Laskar et al., 2004), linking MBT formation to Martian obliquity variations and supporting our multi-stage model.

To constrain the magnitude of late-stage sublimation modification, we compare observed MBT relief with our self-organization simulations. Our results indicate that sublimation must have increased the topographic relief by approximately 3.0 meters ($\theta$ = 45°, $\eta$ = 60%) after the initial self-organization stage, representing the minimum cumulative sublimation-driven modification since terrain stabilization at ~2.9 ± 0.3 Ma. This multi-meter scale is consistent with long-term sublimation rates inferred for shallow subsurface ice on Mars (Hudson et al., 2007; Smith et al., 2009; Dundas et al., 2014, 2015, 2018; Bramson et al., 2017; Fanara et al., 2020; Diniega et al., 2021) and provides a critical constraint for future quantitative models of sublimation specific to MBT landscapes.

## CONCLUSION

In this study, we establish quantitative morphological parameters (DAF, DS, and DW) to characterize MBT and successfully reproduce its brain-like pattern using a dynamic model of self-organized stone transport. However, the self-organization model alone cannot explain the observed relief discrepancy (with <0.5 m and 3.29 ± 0.65 m, respectively). We therefore propose a multi-stage formation mechanism—initial patterning by freeze-thaw cycles followed by late-stage sculpting via sublimation—which implies a paleoclimatic transition from warmer, wetter conditions to colder, drier conditions. The quantified magnitude of late-stage sublimation (~3 m) provides a critical constraint for this landscape, though our results constrain cumulative modification rather than specific process parameters (e.g., rate, critical conditions, substrate dependence). Collectively, these findings demonstrate that MBT serves as a key archive for understanding paleoclimatic conditions and ice/water activity in the mid-latitudes, providing critical context for future missions targeting water ice detection, habitability studies, and paleoclimate reconstruction on Mars.

## ACKNOWLEDGEMENTS

This work is supported by the National Natural Science Foundation of China (Grants No. 42441810 and 42204178). The authors thank MATLAB, which was utilized for morphological parameter extraction. We are grateful for the use of COMSOL Multiphysics (version 6.2), specifically its Phase Field module, for model computation. We also thank ArcGIS (version 10.8) used in geospatial data processing, and the HiRISE team at the University of Arizona for providing the imagery and DTM data for topographic context. The MATLAB code for geometric feature extraction and result visualization is available at https://github.com/zhanglei911/BrainTerrainOnMars. Data associated with this research are available and can be obtained by contacting the corresponding author.

## REFERENCES CITED

Berthling, I., Eiken, T., and Sollid, J.L., 2001, Frost heave and thaw consolidation of ploughing boulders in a mid-alpine environment, Finse, Southern Norway: Permafrost and Periglacial Processes, v. 12, p. 165–177, https://doi.org/10.1002/ppp.367.

Bina, A., and Osinski, G.R., 2021, Decameter-scale rimmed depressions in Utopia Planitia: Insight into the glacial and periglacial history of Mars: Planetary and Space Science, v. 204, p. 105253, https://doi.org/10.1016/j.pss.2021.105253.

Bramson, A.M., Byrne, S., and Bapst, J., 2017, Preservation of Midlatitude Ice Sheets on Mars:

Journal of Geophysical Research: Planets, v. 122, p. 2250–2266, https://doi.org/10.1002/2017JE005357.

Cheng, R.-L., He, H., Michalski, J.R., Li, Y.-L., and Li, L., 2021, Brain-terrain-like features in the Qaidam Basin: Implications for various morphological features on Mars: Icarus, v. 363, p. 114434, https://doi.org/10.1016/j.icarus.2021.114434.

Diniega, S. et al., 2021, Modern Mars' geomorphological activity, driven by wind, frost, and gravity: Geomorphology, v. 380, p. 107627, https://doi.org/10.1016/j.geomorph.2021.107627.

Driver, G., El-Maarry, M.R., Hubbard, B., and Brough, S., 2024, Large Glacier-Like Forms on Mars: Insights From Crater Morphologies and Crater Retention Ages: Journal of Geophysical Research: Planets, v. 129, p. e2023JE008207, https://doi.org/10.1029/2023JE008207.

Dundas, C.M. et al., 2018, Exposed subsurface ice sheets in the Martian mid-latitudes: Science, v. 359, p. 199–201, https://doi.org/10.1126/science.aao1619.

Dundas, C.M., Byrne, S., and McEwen, A.S., 2015, Modeling the development of martian sublimation thermokarst landforms: Icarus, v. 262, p. 154–169, https://doi.org/10.1016/j.icarus.2015.07.033.

Dundas, C.M., Byrne, S., McEwen, A.S., Mellon, M.T., Kennedy, M.R., Daubar, I.J., and Saper, L., 2014, HiRISE observations of new impact craters exposing Martian ground ice: Journal of Geophysical Research: Planets, v. 119, p. 109–127, https://doi.org/10.1002/2013JE004482.

Fanara, L., Gwinner, K., Hauber, E., and Oberst, J., 2020, Present-day erosion rate of north polar

scarps on Mars due to active mass wasting: Icarus, v. 342, p. 113434, https://doi.org/10.1016/j.icarus.2019.113434.

Gallagher, C., Butcher, F.E.G., Balme, M., Smith, I., and Arnold, N., 2021, Landforms indicative of regional warm based glaciation, Phlegra Montes, Mars: Icarus, v. 355, p. 114173, https://doi.org/10.1016/j.icarus.2020.114173.

Hauber, E., Van Gasselt, S., Chapman, M.G., and Neukum, G., 2008, Geomorphic evidence for former lobate debris aprons at low latitudes on Mars: Indicators of the Martian paleoclimate: Journal of Geophysical Research: Planets, v. 113, p. 2007JE002897, https://doi.org/10.1029/2007JE002897.

Hibbard, M., Osinski, G.R., Godin, E., and Kukko, A., 2020, TERRESTRIAL BRAIN TERRAIN AND THE IMPLICATIONS FOR MARTIAN MID-LATITUDES: Ushuaia, Tierra del Fuego, Argentina, The Seventh International Conference on Mars Polar Science and Exploration, Abstract #6023. https://www.hou.usra.edu/meetings/marspolar2020/pdf/6023.pdf.

Hudson, T.L., Aharonson, O., Schorghofer, N., Farmer, C.B., Hecht, M.H., and Bridges, N.T., 2007, Water vapor diffusion in Mars subsurface environments: Journal of Geophysical Research: Planets, v. 112, p. 2006JE002815, https://doi.org/10.1029/2006JE002815.

Kessler, M.A., and Werner, B.T., 2003, Self-Organization of Sorted Patterned Ground: Science, v. 299, p. 380–383, https://doi.org/10.1126/science.1077309.

Laskar, J., Correia, A.C.M., Gastineau, M., Joutel, F., Levrard, B., and Robutel, P., 2004, Long term evolution and chaotic diffusion of the insolation quantities of Mars: Icarus, v. 170, p. 343–364, https://doi.org/10.1016/j.icarus.2004.04.005.

Levy, J.S., Fassett, C.I., Head, J.W., Schwartz, C., and Watters, J.L., 2014, Sequestered glacial ice contribution to the global Martian water budget: Geometric constraints on the volume of remnant, midlatitude debris-covered glaciers: Buried martian glaciers: Journal of Geophysical Research: Planets, v. 119, p. 2188–2196, https://doi.org/10.1002/2014JE004685.

Levy, J., Head, J.W., and Marchant, D.R., 2010, Concentric crater fill in the northern mid-latitudes of Mars: Formation processes and relationships to similar landforms of glacial origin: Icarus, v. 209, p. 390–404, https://doi.org/10.1016/j.icarus.2010.03.036.

Levy, J.S., Head, J.W., and Marchant, D.R., 2009, Concentric crater fill in Utopia Planitia: History and interaction between glacial "brain terrain" and periglacial mantle processes: Icarus, v. 202, p. 462–476, https://doi.org/10.1016/j.icarus.2009.02.018.

Li, A. et al., 2021, Ice needles weave patterns of stones in freezing landscapes: Proceedings of the National Academy of Sciences, v. 118, p. e2110670118, https://doi.org/10.1073/pnas.2110670118.

Li, A., Matsuoka, N., and Niu, F., 2018, Frost sorting on slopes by needle ice: A laboratory simulation on the effect of slope gradient: Earth Surface Processes and Landforms, v. 43, p. 685–694, https://doi.org/10.1002/esp.4276.

Liu, Q.-X., Doelman, A., Rottschäfer, V., De Jager, M., Herman, P.M.J., Rietkerk, M., and Van De Koppel, J., 2013, Phase separation explains a new class of self-organized spatial patterns in ecological systems: Proceedings of the National Academy of Sciences, v. 110, p. 11905–11910, https://doi.org/10.1073/pnas.1222339110.

Mackay, J.R., and Mathews, W.H., 1974, Movement of Sorted Stripes, the Cinder Cone, Garibaldi Park, B. C., Canada: Arctic and Alpine Research, v. 6, p. 347–359, https://doi.org/10.1080/00040851.1974.12003794.

Mangold, N., 2003, Geomorphic analysis of lobate debris aprons on Mars at Mars Orbiter Camera scale: Evidence for ice sublimation initiated by fractures: Journal of Geophysical Research: Planets, v. 108, p. 2002JE001885, https://doi.org/10.1029/2002JE001885.

Matsuoka, N., 2023, How can needle ice transport large stones? Twenty-one years of field observations: Earth Surface Processes and Landforms, v. 48, p. 3115–3127, https://doi.org/10.1002/esp.5685.

Matsuoka, N., Abe, M., and Ijiri, M., 2003, Differential frost heave and sorted patterned ground: field measurements and a laboratory experiment: Geomorphology, v. 52, p. 73–85, https://doi.org/10.1016/S0169-555X(02)00249-0.

Moore, J.M. et al., 2017, Sublimation as a landform-shaping process on Pluto: Icarus, v. 287, p. 320–333, https://doi.org/10.1016/j.icarus.2016.08.025.

Morgan, A.M., Noe Dobrea, E.Z., Pearson, K.A., and Altinok, A., 2025, Crater Retention Timescales

of Brain Coral Terrain Record Past Climatic Change on Mars: The Planetary Science Journal, v. 6, p. 200, https://doi.org/10.3847/PSJ/adf1aa.

Pearson, K.A., Noe, E., Zhao, D., Altinok, A., and Morgan, A.M., 2024, Mapping “Brain Terrain” Regions on Mars Using Deep Learning: The Planetary Science Journal, v. 5, p. 167, https://doi.org/10.3847/PSJ/ad5673.

Smith, P.H. et al., 2009, $H_2 O$ at the Phoenix Landing Site: Science, v. 325, p. 58–61, https://doi.org/10.1126/science.1172339.

Smith, D.E. et al., 2001, Mars Orbiter Laser Altimeter: Experiment summary after the first year of global mapping of Mars: Journal of Geophysical Research: Planets, v. 106, p. 23689–23722, https://doi.org/10.1029/2000JE001364.

Souness, C., Hubbard, B., Milliken, R.E., and Quincey, D., 2012, An inventory and population-scale analysis of martian glacier-like forms: Icarus, v. 217, p. 243–255, https://doi.org/10.1016/j.icarus.2011.10.020.

Wang, C., Chen, J., Chen, L., Sun, Y., Xie, Z., Yin, G., Liu, M., and Li, A., 2022, Experimental and Modeling of Residual Deformation of Soil–Rock Mixture under Freeze–Thaw Cycles: Applied Sciences, v. 12, p. 8224, https://doi.org/10.3390/app12168224.

Yamagishi, C., and Matsuoka, N., 2015, Laboratory frost sorting by needle ice: a pilot experiment on the effects of stone size and extent of surface stone cover: Earth Surface Processes and Landforms, v. 40, p. 502–511, https://doi.org/10.1002/esp.3653.

Zhang, L. et al., 2023, Buried palaeo-polygonal terrain detected underneath Utopia Planitia on Mars by the Zhurong radar: Nature Astronomy, v. 8, p. 69–76, https://doi.org/10.1038/s41550-023-02117-3.

[1]Supplementary material. The details on MBT distribution, morphological parameter extraction, self-organization modeling, and model calibration.